\documentclass[letterpaper,onecolumn,11pt,accepted=2024-11-25]{quantumarticle}
\pdfoutput=1

\usepackage[left=2cm,top=2cm,right=2cm,nohead]{geometry}
\usepackage{amsmath,amssymb,amsthm}
\usepackage{color}
\usepackage{graphicx}
\usepackage[ruled]{algorithm2e}
\usepackage{multirow}
\usepackage{cite}
\usepackage[hidelinks]{hyperref}
\usepackage{url}

\newtheorem{theorem}{Theorem}

\newcommand{\sfrac}[2]{{\ensuremath{\textstyle\frac{#1}{#2}}}}
\newcommand{\ket}[1]{\ensuremath{\vert{#1}\rangle}}
\newcommand{\bra}[1]{\ensuremath{\langle{#1}\vert}}

\newcommand{\half}[0]{\sfrac{1}{2}}

\newcommand{\Tr}{\ensuremath{\mathrm{Tr}}}
\newcommand{\Cop}{{O}}
\newcommand{\Coper}{{\mathcal{O}}}
\newcommand{\Pauli}[1]{\ensuremath{{{#1}}}}

\usepackage[normalem]{ulem}

\begin{document}

\title{Techniques for learning sparse Pauli-Lindblad noise models}
\author{Ewout van den Berg}
\orcid{0000-0002-0991-3397}
\author{Pawel Wocjan}
\orcid{0000-0000-0000-0000}
\affiliation{IBM Quantum, IBM Thomas J.~Watson Research Center, Yorktown
  Heights, NY, USA}

\begin{abstract}
  Error-mitigation techniques such as probabilistic error cancellation
  and zero-noise extrapolation benefit from accurate noise models. The
  sparse Pauli-Lindblad noise model is one of the most successful
  models for those applications. In existing implementations, the
  model decomposes into a series of simple Pauli channels with one-
  and two-local terms that follow the qubit topology. While the model
  has been shown to accurately capture the noise in contemporary
  superconducting quantum processors for error mitigation, it is
  important to consider higher-weight terms and effects beyond
  nearest-neighbor interactions.  For such extended models to remain
  practical, however, we need to ensure that they can be learned
  efficiently. In this work we present new techniques that accomplish
  exactly this. We introduce twirling based on Pauli rotations, which
  enables us to automatically generate single-qubit learning
  correction sequences and reduce the number of unique fidelities that
  need to be learned. In addition, we propose a basis-selection
  strategy that leverages graph coloring and uniform covering arrays
  to minimize the number of learning bases. Taken together, these
  techniques ensure that the learning of the extended noise models
  remains efficient, despite their increased complexity.
\end{abstract}

\section{Introduction}

An important primitive in many quantum algorithms, such as variational
quantum eigensolvers~\cite{PER2014MSYa,MCC2016RBAa,TIL2022CCPa}, is
the accurate estimation of quantum observable expectation
values. Until fault tolerance is achieved, these estimates will be
affected by noise inherent in current quantum processors and the
associated control systems. A practical way of reducing the effect of
noise is the use of error-mitigation techniques. These techniques can
be roughly divided into two complementary groups, depending on what
part of the quantum circuit execution is considered. The first group
of mitigation algorithms is aimed specifically at reducing
state-preparation and measurement
errors~\cite{BER2022MTa,BRA2021SKMa,HIC2021BNa}. The second group
mostly considers noise associated with the application of gates, and
contains techniques such as zero-noise
extrapolation~\cite{TEM2017BGa,LI2017Ba,KAN2019TCMa} and probabilistic
error cancellation~\cite{TEM2017BGa}. In zero-noise extrapolation
(ZNE), observable estimates are obtained at carefully controlled noise
levels, which are then extrapolated to the point were gates are
presumed noiseless. Meanwhile, in probabilistic error cancellation
(PEC), noiseless operators are implemented as (quasi)-probabilistic
mixtures of potentially noisy gates. This technique is leveraged
in~\cite{BER2023MKTa} to mitigate noise associated with Hermitian
Clifford operators by learning a sparse Pauli-Lindblad noise model
(discussed in more detail in the next section), and subsequently
applying the inverse noise map in a quasi-probabilistic manner to
obtain unbiased estimates of the expectation value of desired
observables. When noise models associated with the gates in a circuit
are available they provide an ideal way of controlling the noise level
in the context of
ZNE~\cite{END2018BLa,MAR2021SZa,FER2024HVNa,MCD2022MSSa}. Indeed,
noise amplification based on the sparse Pauli-Lindblad noise model was
used to great effect in this manner to mitigate noise in a 127-qubit
Ising model in~\cite{KIM2023EAWa}.  Aside from these applications,
noise models can help characterize the performance of quantum
processors and identify possible points of improvements.

In this work, we focus on techniques for learning extended versions of
the sparse Pauli-Lindblad noise model proposed in~\cite{BER2023MKTa}
and reviewed in Section~\ref{Sec:NoiseModel}. We introduce
Pauli-rotation twirling in Section~\ref{Sec:Twirling}. In
Section~\ref{Sec:Applications} we show how rotation twirling can be
used in shaping the noise and the design of noise-learning
sequences. Both techniques are leveraged in
Section~\ref{Sec:BasisSelection} to improve learning protocols through
optimized basis selection. We provide concluding remarks in
Section~\ref{Sec:Conclusions}.

\section{Sparse Pauli-Lindblad noise models}\label{Sec:NoiseModel}

Noisy quantum gates $\tilde{\mathcal{U}}$ can be modeled as an ideal
operator $\mathcal{U}$ preceded by a noise channel $\tilde{\Lambda}$.
By restricting $\mathcal{U}$ to represent a Clifford operator, we can
efficiently implement Pauli twirling~\cite{PhysRevLett.76.722,knill2004fault,kern2005quantum,geller2013efficient,wallman2016noise},
which turns noise channels into Pauli channels. We can therefore
assume that $\tilde{\Lambda}$ is a Pauli channel, namely
$\tilde{\Lambda}(\rho) = \sum_i \alpha_i P_i\rho P_i$ where $P_i$
represents a Pauli operator. A general $n$-qubit Pauli channel is
characterized by $4^n$ coefficients, so, clearly, the representation
and learning of the full channel does not scale well in the number of
qubits. Moreover, even if there are only a small number of non-zero
$\alpha_i$ parameters, the inverse noise map generally still requires
the evaluation and inversion of all $4^n$ Pauli
fidelities\footnote{When $\tilde{\Lambda}$ is a Pauli channel, it
  holds that $\tilde{\Lambda}(P_b) = f_bP_b$ for all Pauli operators
  $P_b$. The scalars $f_b$ are therefore commonly referred to as the
  eigenvalues of the channel.}
$f_b = \sfrac{1}{2^n}\Tr(P_b\tilde{\Lambda}(P_b))$.

The Pauli-Lindblad noise model introduced in~\cite{BER2023MKTa} is an
alternative model that is more scalable and comes with convenient
properties. The model is defined as
$\Lambda(\rho) = \exp[\mathcal{L}]\rho$, where
$\mathcal{L}(\rho) = \sum_{k\in\mathcal{K}} \lambda_k(P_k\rho
P_k^{\dag} - \rho)$ is a Lindbladian with Pauli jump operators
$\{P_k\}_{k\in\mathcal{K}}$. It can be equivalently expressed as the
composition of a series of simple Pauli channels:
\[
\Lambda(\rho) = \mathop{\bigcirc}_{k\in\mathcal{K}}\left(
w_k \cdot + (1-w_k)P_k\cdot P_k^{\dag}\right)(\rho),
\]
with $w_k = (1 + e^{-2\lambda_k})/2$ and $\lambda_k \geq 0$. The
expressivity of the model depends on the choice of $\mathcal{K}$, but
always remains a strict subset of all possible Pauli channels. By
restricting $\mathcal{K}$ to the set of all one- and two-local Pauli
terms following the qubit topology, we obtain a sparse
Pauli-Lindblad noise model~\cite{BER2023MKTa}.

Learning of the model parameters
$\lambda = \{\lambda_k\}_{k\in\mathcal{K}}$ makes use of the following
relation to Pauli fidelities:
\begin{equation}\label{Eq:Fb}
f_b
= \sfrac{1}{2^n}\Tr(P_b\Lambda(P_b))
= \exp\left(-2
\sum_{k\in\mathcal{K}}\lambda_k \langle b,k\rangle_{\mathrm{sp}}
\right),
\end{equation}
where $\langle b,k\rangle_{\mathrm{sp}}$ is the symplectic inner
product between Paulis $P_b$ and $P_k$, which has a value of 0 when
the terms commute and 1 otherwise. For a vector
$f = \{f_b\}_{b\in\mathcal{B}}$ of measured fidelities and
binary matrix $M(\mathcal{B}, \mathcal{K})$ with entries
$\langle b,k\rangle_{\mathrm{sp}}$ it follows from Eq.~\eqref{Eq:Fb}
that $M(\mathcal{B}, \mathcal{K})\lambda = -\log(f)/2$, where
$\lambda$ is the vector of model parameters
$\{\lambda_k\}_{k\in\mathcal{K}}$. Learning the noise model parameters
based on a vector of estimated fidelities $\hat{f}$ can therefore be
done using nonnegative least-squares optimization:
\begin{equation}\label{Eq:NNLS}
\mathop{\mathrm{minimize}}_{\lambda \geq 0}\quad
\half\Vert M(\mathcal{B}, \mathcal{K})\lambda + \log(\hat{f})/2\Vert.
\end{equation}
The solution of this problem is unique whenever
$M(\mathcal{B}, \mathcal{K})$ is full column rank, and a convenient
result in this context is the following:
\begin{theorem}[{{\!\!\cite[Theorem SIV.1]{BER2023MKTa}}}]\label{Thm:FullRankM}
  For non-empty support sets $\mathcal{S}_i \subseteq [n]$, denote by
  $\mathcal{P}_i$ all $n$-qubit non-identity Paulis supported on
  $\mathcal{S}_i$, and their union by
  $\mathcal{P} = \bigcup_i \mathcal{P}_i$. Then
  $M(\mathcal{P}, \mathcal{P})$ is full rank.
\end{theorem}
\noindent
From this we conclude that for the above two-local noise model, it
suffices to choose $\mathcal{B} = \mathcal{K}$. Adding more terms to
$\mathcal{B}$ does not affect the column rank, but does result in an
overcomplete set of equations, which can help prevent overfitting when
using noisy fidelity estimates $\hat{f}$.

Estimating the fidelities of Pauli channels can be done using several
protocols, for
instance~\cite{FLA2020Wa,ERH2019WPMa,PhysRevX.4.011050,HEL2019XVWa}.
To understand how these work, note that we can express the initial
quantum state $\rho_0$ as a weighted sum of Pauli terms
$\frac{1}{2^n}\sum_{i}\alpha_i P_i$. In particular, the zero state
$\ket{0}\bra{0}$ can be written as the sum of all Pauli-Z terms, each
with $\alpha_i = 1$.  The Pauli transfer matrix of any Clifford
operator is a signed permutation matrix, while that of a Pauli channel
is a diagonal matrix whose entries correspond to the channel Pauli
fidelities. Since Pauli terms never split into multiple terms under
these transformation, we can track individual Pauli terms and their
associated fidelity as successive noisy Clifford gates are
applied. Repeated application of a noisy Clifford operator
$\tilde{\mathcal{U}}$ therefore leads to a product of fidelities that
depends on the transformation of the Pauli term. Whenever the number
of applications of the operator, or depth, $k$ is such that
$\mathcal{U}^k(P) = P$ we can learn the fidelities free from any
state-preparation and measurement (SPAM)
errors~\cite{ERH2019WPMa,BER2022MTa}. To keep learning manageable, we
must avoid products of many fidelities and therefore require
$\mathcal{U}$ to be Hermitian; that is $\mathcal{U}^2 = I$. More
specifically, we allow $\mathcal{U}$ to be a \emph{layer} of
simultaneous single- and two-qubit Hermitian Clifford gates. For such
layers we can estimate products of at most two unique fidelities, say
$(f_1f_2)^{k/2}$, for any even depth $k$. The resulting fidelities can
then be found by fitting an exponential curve through the fidelity
estimates at different depths~\cite{FLA2020Wa}.

With appropriate basis changes, this method allows us to measure the
fidelity pair $f_if_j$ for any Pauli pair $P_i$ and
$P_j = \mathcal{U}(P_i)$. However, ideally we would measure individual
fidelities, rather than pairs. Whenever the support of $P_i$ and $P_j$
is the same we can interleave applications of $\tilde{\mathcal{U}}$
with appropriate single-qubit correction gates to map $P_j$ back to
$P_i$. For instance, we can change a Pauli $X$ term to a Pauli $Z$
term by applying a single-qubit Hadamard gate. We consider such
correction sequences in Section~\ref{Sec:CorrectionSequences}. Since
the identity is invariant under conjugation by any unitary, it is not
possible to change an identity term to a non-identity term, and vice
versa. In other words, it is not possible to change the support of a
Pauli by means of single-qubit gates, and it was shown
in~\cite{CHE2020YZFa} that fidelity pairs corresponding to Paulis with
different supports cannot be learned in a SPAM-free manner.  In such
cases, we can write $(\alpha f_i)(f_j / \alpha) = f_if_j$ for any
$\alpha \neq 0$ and there is therefore no unique way to extract the
individual fidelities from their product.

As a practical solution, a \emph{symmetry assumption} was introduced
in~\cite{BER2023MKTa}, which takes $f_i$ and $f_j$ to be equal
whenever disambiguation is not possible. With this we can learn $f_i$
by measurements in a Pauli basis that contains $P_i$ or its
corresponding term $P_j$. Although measurements obtained from circuits
in a single fixed basis can be used to estimate exponentially many
fidelities, not all fidelities in our target set $\mathcal{B}$ can be
learned in a single basis. It is therefore natural to consider the
minimal set of bases from which we can learn all desired
fidelities. For the two-local noise model, it was shown
in~\cite{BER2023MKTa} that measurements in nine different bases
suffice for the qubit topologies under consideration.  In
Section~\ref{Sec:BasisSelection} we present an efficient
basis-selection algorithm that applies for our extended noise models
and includes the existing learning algorithm as a special case.

\section{Pauli rotation twirls}\label{Sec:Twirling}

Twirling is a technique that is used to shape noise channels. Pauli
twirling of a noise channel $\tilde{\Lambda}$ associated to a noisy
$n$-qubit Clifford operator
$\tilde{\Coper} = \Coper\circ \tilde{\Lambda}$, with
$\Coper(\rho) = \Cop\rho\Cop^{\dag}$ is given by
\begin{equation}\label{Eq:OriginalTwirl}
\mathbb{E}_{i\in\mathcal{I}} \left[O P_i^{\dag} \tilde{\Lambda}(P_i \rho P_i^{\dag}) P_i O^{\dag}\right],
\end{equation}
where $\mathcal{I}$ represents the $n$-qubit Pauli group, or an
appropriate subset thereof~\cite{cai2019constructing}. Since the noise
of interest appears in the context of a gate, rather than by itself,
we cannot just apply Pauli operators prior to and following the noise
channel. Instead, the circuit implementation of a given twirl element
$P_i$ relies on the identity $OP_i^{\dag} = \sigma_i Q_i^{\dag} O$,
which follows from the $OP_iO^{\dag} = \sigma_i Q_i$ for some Pauli
$Q_i$ with sign $\sigma_i$. Using this identity we can rewrite
Eq.~\eqref{Eq:OriginalTwirl} in terms of the noisy operator
$\tilde{\Coper}$ and obtain the equivalent expression
\[
\mathbb{E}_{i\in\mathcal{I}} \left[
Q_i^{\dag} \tilde{\mathcal{O}}(P_i \rho P_i^{\dag}) Q_i
\right],
\]
which shows that Paul twirling can be implemented in practice.

\subsection{Pauli rotation twirl}\label{Sec:PauliRotationTwirl}

We now generalize the Pauli twirl to a twirl based on Pauli rotation
operators. This twirl will help reduce the number of unique fidelities
and thereby simplify the noise learning process. The rotation operator
$R_{P}(\theta)$ for Pauli operator $P$ and angle $\theta$ is defined
as
\begin{equation}\label{Eq:RP}
R_{\Pauli{P}}(\theta) = \exp\left(-i\sfrac{\theta}{2}\Pauli{P}\right)
=\cos(\theta/2)I - i\sin(\theta/2)P.
\end{equation}
The corresponding twirl, with fixed $\theta$ for simplicity, is then given by
\begin{equation}\label{Eq:RotationTwirl}
\mathbb{E}_{j\in\mathcal{J}} \left[
O R_{P_j}(-\theta)\tilde{\Lambda}\big(R_{P_j}(\theta)\rho R_{P_j}(-\theta)\big) R_{P_j}(\theta)\Cop^{\dag}
\right],
\end{equation}
where $\mathcal{J}$ represents (a subset of) the Pauli group. Note
that setting $\theta=\pi$ in Eq.~\eqref{Eq:RP} yields the identity
$R_P(\pi) = (-i)P$. Therefore, applying the rotation twirl with
$\theta=\pi$ reduces to the standard Pauli twirl. For the
implementation of the twirl, we again need to conjugate the twirl
operators by $O$, which satisfies
\begin{equation}\label{Eq:ORpO=Rq}
\Cop R_{P_j}(\theta) \Cop^{\dag} =
\Cop \exp\left(-i\sfrac{\theta}{2}P_j\right) \Cop^{\dag} =
\exp\left(-i\sfrac{\theta}{2}\Cop P_j\Cop^{\dag}\right) =
\exp\left(-i\sigma_j\sfrac{\theta}{2}Q_j\right)
= R_{Q_j}(\sigma_j\theta).
\end{equation}
Post-multiplying by $\Cop$ then gives
$OR_{P_j}(\theta) = R_{Q_j}(\sigma_i\theta) O$, which allows us to
rewrite Eq.~\eqref{Eq:RotationTwirl} as
\begin{equation}
\mathbb{E}_{j\in\mathcal{J}} \left[
R_{Q_j}(-\sigma_j\theta)\ \tilde{\Coper}\big(R_{P_j}(\theta)\rho R_{P_j}(-\theta)\big)\ R_{Q_j}(\sigma_j \theta)
\right].
\end{equation}
Although the twirl is now expressed in terms of the noisy operator
$\tilde{\Coper}$, implementing general rotations for arbitrary Pauli
terms $P_j$ and angles $\theta$ may require non-trivial circuits that
contain two-qubit gates, which themselves may be noisy. We may
therefore want to limit $\mathcal{J}$ to a subset of Pauli operators
for which the $P_j$ and $Q_j$ rotation operations can be implemented
using single-qubit rotations, which are readily implemented on
contemporary quantum processors. Since Pauli operators decouple into
single-qubit Pauli terms, this holds trivially for all $\mathcal{J}$
in the special case of Pauli twirls. For all other values of $\theta$,
it suffices to find values of $j$ for which both $P_j$ and $Q_j$ have
weight one. For particular angles $\theta$ we can then further
simplify the implementation and express the rotation in terms of basic
single-qubit gates, as summarized in
Table~\ref{Table:RotationOperators}.

We now show that for each two-qubit Hermitian Clifford gate $O$, there
always exists a two-qubit Pauli of the form $PI = P\otimes I$ such
that its corresponding Pauli $\sigma O(PI)O^\dagger$ is also
weight-one, and likewise for a Pauli of the form $IP$. This
demonstrates that non-trivial Pauli rotation twirling exists for these
gates for any angle $\theta$.  To show that a two-qubit Pauli of the
form $PI$ exists, suppose by contradiction that \Pauli{XI},
\Pauli{YI}, and \Pauli{ZI} all map to weight-two Paulis.  Because
conjugation by a Clifford operator preserves commutation relations,
these new Paulis mutually anticommute and must therefore be of the
form $\Pauli{S}\times\{\Pauli{X},\Pauli{Y},\Pauli{Z}\}$ or
$\{\Pauli{X},\Pauli{Y},\Pauli{Z}\}\times \Pauli{S}$.  Any remaining
weight-two Pauli operator can be verified to anticommute with one of
these new operators. Since Pauli operators \Pauli{IX}, \Pauli{IY}, and
\Pauli{IZ} commute with all Pauli operators of the form \Pauli{PI},
they therefore cannot map to any weight-two Pauli, and must retain
their original support. But even with the same support at least one
will anticommute with the image of \Pauli{XI}, \Pauli{YI}, and
\Pauli{ZI}, which contradicts the definition of Clifford
operators. Using an analogous derivation, we can show a similar result
for Paulis of the form $IP$.

\begin{table}[t]
\centering
\begin{tabular}{lccc}
\hline
Rotation & $\theta=\pi$ & $\theta=\pi/2$ & $\theta=\pi/4$\\
\hline
$R_{\Pauli{X}}(\theta)$
& $(-i)X$
& $\left(\sfrac{1-1j}{\sqrt{2}}\right)S_X$
& $\left(e^{-i\pi/8}\right)HTH$
\\
$R_{\Pauli{Y}}(\theta)$
& $(-i)Y$
& $HZ$
& $\left(e^{-i\pi/8}\right)S_X^{\dag}TS_X$
\\
$R_{\Pauli{Z}}(\theta)$
& $(-i)Z$
& $\left(\sfrac{1-1j}{\sqrt{2}}\right)S$
& $\left(e^{-i\pi/8}\right)T$
\\
\hline
\end{tabular}
\caption{Expression of single-qubit Pauli rotations in terms of basic
  gates for three specific $\theta$ values. The operations corresponding
to rotations by $-\theta$ are given by the adjoint of the listed
operations. For instance, $R_{\Pauli{Z}}(-\pi/4) =
(e^{i\pi/8})T^{\dag}$ and $R_{\Pauli{Y}}(-\pi/2) = ZH$.}\label{Table:RotationOperators}
\end{table}

\subsection{Classes of two-qubit Hermitian Clifford operators}\label{Sec:Classes}

For the remainder of the paper it will be convenient to have a clear
characterization of the changes in the support of Pauli operators
following conjugation by two-qubit Hermitian Clifford operators.  In
particular, this classification enables us to determine suitable
weight-one Pauli pairs, thereby extending the existence proof from
Section~\ref{Sec:PauliRotationTwirl}.

Transitions in support happen when an $\Pauli{I}$ term
changes to an $\{\Pauli{X},\Pauli{Y},\Pauli{Z}\}$ term, or vice
versa. As an example, consider the conditional-Z (CZ) gate. It maps $\Pauli{XI}$ to
$\Pauli{XZ}$ changing the support, whereas it maps
$\Pauli{XX}$ to $\Pauli{YY}$ maintaining the
original support.

\begin{figure}[t!]
\centering
\begin{tabular}{cccc}
Class 1 & Class 2 & Class 3 & Class 4\\[5pt]
\includegraphics[width=0.224\textwidth]{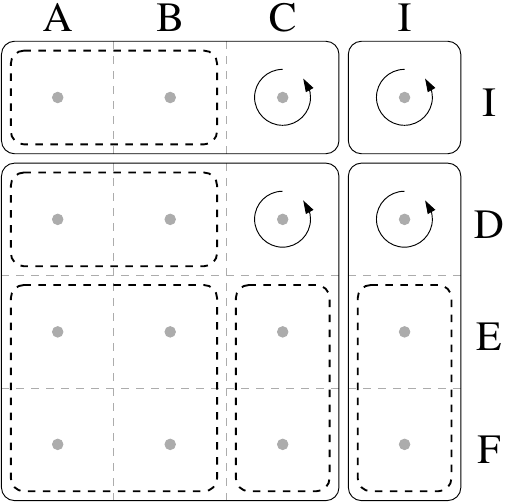}&
\includegraphics[width=0.224\textwidth]{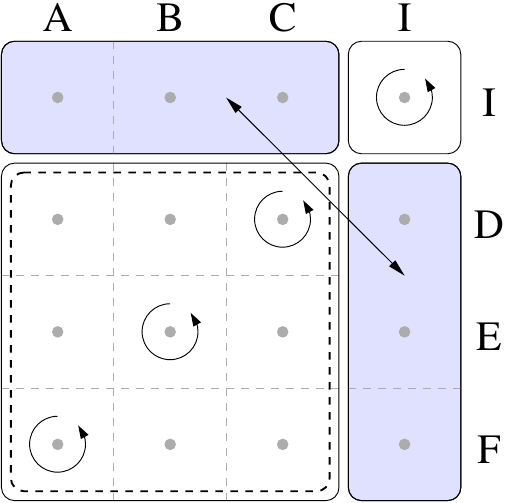}&
\includegraphics[width=0.224\textwidth]{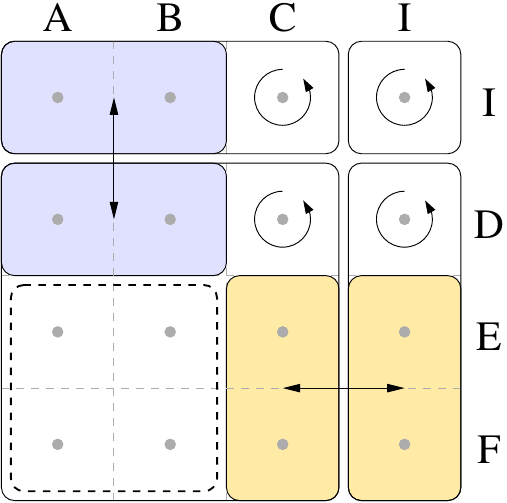}&
\includegraphics[width=0.224\textwidth]{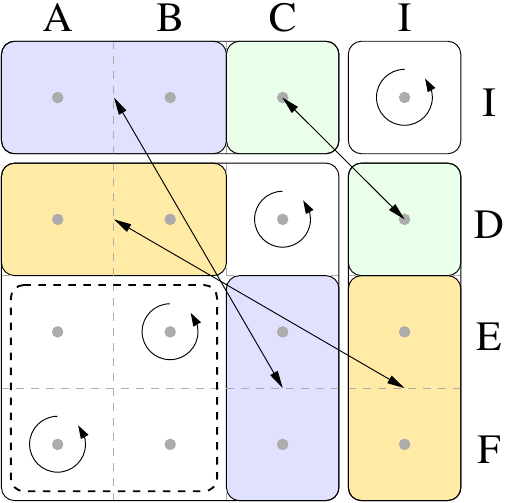}
\end{tabular}
\caption{Pauli support transitions for four classes of two-qubit
  Hermitian Clifford operators, where
  $\{\Pauli{A},\Pauli{B},\Pauli{C}\}$ and
  $\{\Pauli{D},\Pauli{E},\Pauli{F}\}$ are arbitrary permutations of
  $\{\Pauli{X},\Pauli{Y},\Pauli{Z}\}$. Column (row) labels correspond
  to the Pauli on the first (second) qubit. Pauli terms within each of
  the dashed regions remain in those regions after conjugation. The
  cyclic arrow indicates mapping of a Pauli to itself.}\label{Fig:Classes}
\end{figure}

We show that Hermitian two-qubit Clifford operators can be partitioned
into four classes depending on how their mapping affects the support
of Pauli operators. These classes are illustrated in
Figure~\ref{Fig:Classes}, which groups Paulis of the form
$\{\Pauli{I},\Pauli{A},\Pauli{B},\Pauli{C}\}\otimes\{\Pauli{I},\Pauli{D},\Pauli{E},\Pauli{F}\}$
in boxes, based on their support, where
$\{\Pauli{A},\Pauli{B},\Pauli{C}\}$ and
$\{\Pauli{D},\Pauli{E},\Pauli{F}\}$ represent arbitrary permutations
of $\{\Pauli{X},\Pauli{Y},\Pauli{Z}\}$. We will use three basic
properties:

\begin{enumerate}
\item For any Clifford operator $\Cop$, the
conjugation of the product of two Paulis is equal to the product of
their individual conjugations:
$\Cop(\Pauli{P}\cdot\Pauli{Q})\Cop^{\dag}
= \Cop\Pauli{P}(\Cop^{\dag}\Cop)\Pauli{Q}\Cop^{\dag}
= (\Cop\Pauli{P}\Cop^{\dag})(\Cop\Pauli{Q}\Cop^{\dag})$.

\item Commutation relations between Paulis are
preserved under conjugation by Clifford operators: if Paulis
$\Pauli{P}$ and $\Pauli{Q}$ commute then so do
$\Cop\Pauli{P}\Cop^{\dag}$ and $\Cop\Pauli{Q}\Cop^{\dag}$.

\item Hermiticity of $\Cop$ implies that
whenever $\Pauli{P}$ maps to $\Pauli{Q} = \Cop\Pauli{P}\Cop^{\dag}$
then $\Pauli{Q}$ maps to $\Pauli{P}$.
\end{enumerate}

\noindent With this, we can characterize the four different
classes. Throughout the analysis we ignore phase factors in
expressions such as $\Pauli{X}\cdot\Pauli{Y} = \Pauli{Z}$.

\paragraph{Class 1}
If $\Pauli{AI}$ and $\Pauli{BI}$ map to Paulis of the form
$\Pauli{PI}$ and $\Pauli{QI}$, then it follows from property 1 that
$\Pauli{CI} = (\Pauli{AI})\cdot (\Pauli{BI})$ maps to
$(\Pauli{P}\cdot\Pauli{Q})\Pauli{I}$ and therefore also does not
change support. It further follows from property 2 that $\Pauli{ID}$
cannot change support, otherwise it would anticommute with one of
$\Pauli{PI}$, $\Pauli{QI}$, or
$(\Pauli{P}\cdot\Pauli{Q})\Pauli{I}$. The same applies to $\Pauli{IE}$
and $\Pauli{IF}$. This setting occurs whenever $\Cop$ decouples into
two single-qubit operators, i.e., $\Cop= \Cop_1\otimes \Cop_2$. At
least one Pauli in each of the top and right weight-one boxes must map
to itself; cyclic changes between three terms would imply that
$\Cop^2 \neq \Pauli{I}$ and therefore contradict Hermiticity of
$\Cop$.

\paragraph{Class 2}

Similar to class 1, suppose that $\Pauli{AI}$ and $\Pauli{BI}$
respectively map to $\Pauli{IF}$ and $\Pauli{IE}$, then it follows
from property 1 that $\Pauli{CI}$ must map to $\Pauli{ID}$, and that
all weight-two terms retain their support. Using properties 1 and 3 we
find that $\Pauli{AF} = (\Pauli{AI})\cdot (\Pauli{IF})$ maps to
$ (\Pauli{IF})\cdot (\Pauli{AI}) = \Pauli{AF}$. Applying the same
reasoning to the other pairs, we find that $\Pauli{BE}$ and
$\Pauli{CD}$ also map to themselves.

\paragraph{Class 3}
Suppose $\Pauli{CI}$ maps to $\Pauli{PI}$ and $\Pauli{AI}$ maps to
$\Pauli{ST}$, then, using property 1, we find that $\Pauli{BI}$ must
map to $(\Pauli{S}\cdot \Pauli{P})\Pauli{T}$, and therefore also
changes support. Hermitian Clifford operators map one Pauli to another
and vice versa. Since $\Pauli{AI}$ and $\Pauli{BI}$ already map to
other Pauli terms, we conclude that $\Pauli{CI}$ must map to itself,
that is, $P=C$. In order for \Pauli{ST} and
$(\Pauli{S}\cdot\Pauli{P})\Pauli{T}$ to anticommute with $\Pauli{CI}$,
both $\Pauli{S}$ and $\Pauli{S\cdot P}$ must differ from $\Pauli{C}$,
which means that $\{\Pauli{AI}, \Pauli{BI}\}$ maps to
$\{\Pauli{AT}, \Pauli{BT}\}$ in some order.  The terms in the
bottom-right box (\Pauli{ID}, \Pauli{IE}, and \Pauli{IF}) must map to
Paulis that commute with $\Pauli{AT}$ and $\Pauli{BT}$. This excludes
terms in the $3\times 3$ weight-two box that are in the same row or
column. This leaves exactly two feasible elements and exactly one of
$\Pauli{ID}$, $\Pauli{IE}$, $\Pauli{IF}$ must therefore retain its
support (if multiple elements retain their support, class 1 would
apply). Assume this element is $\Pauli{ID}$, then following the same
argument as for $\Pauli{CI}$, we conclude that $\Pauli{ID}$ must map
to itself. Since both $\Pauli{CI}$ and $\Pauli{ID}$ map to themselves,
so does their product $\Pauli{CD}$. Finally, note that the unspecified
term $\Pauli{T}$ above must be equal to $\Pauli{D}$, since that row is
excluded for the two terms from the bottom-right box that change
support, as illustrated in Figure~\ref{Fig:Classes}.

\paragraph{Class 4}
Suppose $\Pauli{CI}$ maps to $\Pauli{ID}$ and that $\Pauli{AI}$ maps
to weight-two Pauli $\Pauli{PQ}$. Following property 1,
$\Pauli{BI} = (\Pauli{AI})\cdot(\Pauli{CI})$ maps to
$\Pauli{P}(\Pauli{Q}\cdot\Pauli{D})$, and
$\Pauli{CD} = (\Pauli{CI})\cdot(\Pauli{ID})$ maps to itself. From
property 2 we further have that $\Pauli{PQ}$ and
$\Pauli{P}(\Pauli{Q}\cdot\Pauli{D})$ must anticommute with
$\Pauli{ID}$, which means that both $\Pauli{Q}$ and
$\Pauli{Q}\cdot\Pauli{D}$ both differ from $\Pauli{D}$, so without
loss of generality we can set $\Pauli{Q}=\Pauli{E}$.
Since $\Pauli{ID}$ and $\Pauli{AI}$ commute, so do their images
$\Pauli{CI}$ and $\Pauli{PQ}$, which implies that
$\Pauli{P}=\Pauli{C}$.
We conclude that $\Pauli{AI}$ maps to $\Pauli{PQ} =
\Pauli{CE}$. Applying properties 1 and 3 on both terms we find that
$(\Pauli{AI})\cdot(\Pauli{CE}) = \Pauli{BE}$ maps to itself, and
likewise for $\Pauli{AF}$ through multiplication by $\Pauli{CD}$.
Finally, combining the mappings for $\Pauli{CI}$, $\Pauli{CE}$, and
$\Pauli{ID}$ we find that $\Pauli{IE}$ and $\Pauli{IF}$ respectively
map to $\Pauli{AD}$ and $\Pauli{BD}$.

\paragraph{Excluding other classes}
The only possible class remaining is one in which all weight-one Pauli
terms in the top box map to weight-two Paulis. In order for the
resulting Paulis to satisfy property 2, they should all mutually
anticommute and therefore lie in a single row or column of the
$3\times 3$ weight-two box. As a result, none of the remaining
elements in the box commutes with all three of these Paulis.  In order
to satisfy property 2, the weight-one Pauli terms in the right box
must therefore all map to Paulis within the same box. But this would
imply case 1 and therefore leads to a contradiction. We conclude that
the four classes discussed above capture all weight-two Hermitian
Clifford operators.

\section{Applications of the Pauli rotation twirl}\label{Sec:Applications}

In preparation for basis selection, discussed in
Section~\ref{Sec:BasisSelection}, we now consider two applications of
the Pauli rotation twirl. The first application shapes Pauli channels
by averaging certain fidelities and relies on nested twirling with
different twirl sets $\mathcal{J}$. Doing so reduces the number of
unique Pauli fidelities that need to be learned and allows us to
reduce the number of measurement bases used to learn the sparse
Pauli-Lindblad noise model. The second application concerns the design
of single-qubit correction sequences in learning circuits to enable
extraction of individual channel fidelities. In both applications we
fix $\theta = \pi/2$.

\subsection{Rotation twirls for fidelity averaging}\label{Sec:ExtraTwirling}

\begin{figure}
\centering\setlength{\tabcolsep}{12pt}
\begin{tabular}{ccc}
\raisebox{-50pt}{\includegraphics[height=115pt]{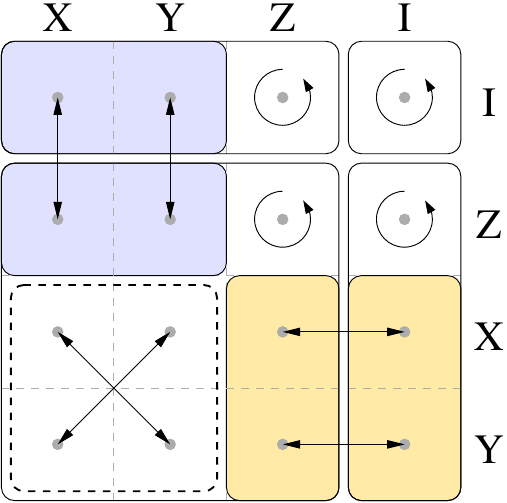}}
&
\setlength{\tabcolsep}{8pt}
\begin{tabular}{c}
\includegraphics[width=0.115\textwidth]{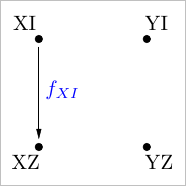}\\[12pt]
\includegraphics[width=0.115\textwidth]{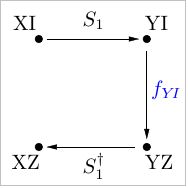}
\end{tabular}
&
\setlength{\tabcolsep}{8pt}
\begin{tabular}{cc}
\includegraphics[width=0.115\textwidth]{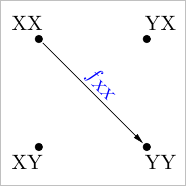}&
\includegraphics[width=0.115\textwidth]{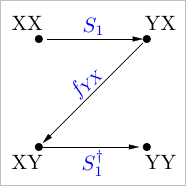}\\[12pt]
\includegraphics[width=0.115\textwidth]{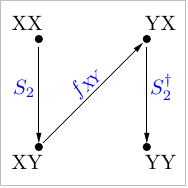}&
\includegraphics[width=0.115\textwidth]{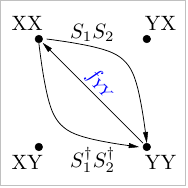}
\end{tabular}
\\
\\[-8pt]
({\bf{a}}) & ({\bf{b}}) & ({\bf{c}})
\end{tabular}
\caption{Illustration of ({\bf{a}}) the conjugation of Pauli terms under
  class-3 operator CZ; ({\bf{b}}) two trajectories of moving from
  $\Pauli{XI}$ to $\Pauli{XZ}$ when applying a phase twirl $\{S_1, I_1\}$
  on qubit one. The effective transition fidelity resulting from the
  twirl is given by the average fidelity
  $(f_{\Pauli{XI}} + f_{\Pauli{YI}})/2$; ({\bf{c}}) four trajectories of moving from $\Pauli{XX}$ to $\Pauli{YY}$
  when applying phase twirls on both qubits. Here too the effective transition fidelity is
  given by the average of the path fidelities, assuming that single-qubit gates are noiseless.}\label{Fig:CZGate}
\end{figure}

As seen from Table~\ref{Table:RotationOperators}, weight-one Pauli
$X$, $Y$, and $Z$ rotations with angle $\theta = \pi/2$ result
respectively in $S_X$, $HZ$ and $S$ operations, where $S_X$ is the
square root of $\Pauli{X}$, $H$ is the Hadamard gate, and $S$ is the
phase gate. These operations are easy to implement and have the
special property that their application permutes single-qubit Pauli
terms. In particular, ignoring signs, conjugation with $HZ$ exchanges
Pauli $\Pauli{X}$ and $\Pauli{Z}$ terms, $S$ exchanges $\Pauli{X}$ and
$\Pauli{Y}$, and $S_X$ exchanges $\Pauli{Y}$ and $\Pauli{Z}$ terms.

We now explain how this property can be used to average certain
fidelities in a noisy CZ gate, where it is assumed that the noise
channel $\tilde{\Lambda}$ preceding the ideal CZ gate has already been
shaped to a Pauli channel. For reference, we illustrate the Pauli
transitions for the CZ gate in Figure~\ref{Fig:CZGate}(a).

Choosing a twirl set $\mathcal{J}_1 = \{\Pauli{II}, \Pauli{ZI}\}$
amounts to twirling the noisy CZ gate by Pauli rotations
$\{I, R_Z(\sfrac{\pi}{2})\}$ on the first qubit. The operation
resulting from the twirl is given by the expectation in
Eq.~\eqref{Eq:RotationTwirl}, which is taken over the two values
$j\in\mathcal{J}_1$.  For $j = \Pauli{II}$, all rotations reduce to
the identity and the noise channel $\tilde{\Lambda}$ applies directly
to $\Pauli{XI}$, which results in fidelity of
$f_\Pauli{XI}$. Application of the ideal CZ gate then changes
$\Pauli{XI}$ to $\Pauli{XZ}$. On the other hand, for $j=\Pauli{ZI}$,
we start by applying a Pauli-Z rotation, which is equivalent to a
phase gate and therefore changes $\Pauli{XI}$ to $\Pauli{YI}$. The
noise channel then incurs a fidelity term $f_{\Pauli{YI}}$, but leaves
the Pauli term itself unchanged. The ideal CZ gate and the second
rotation gate that follows then successively map $\Pauli{YI}$ to
$\Pauli{YZ}$ and $\Pauli{XZ}$, which is the same final Pauli term as
before (see the illustration in Figure~\ref{Fig:CZGate}(b)). To see
why, observe that Eq.~\eqref{Eq:ORpO=Rq} can be
rewritten as
\begin{equation}\label{Eq:C=RQCRP}
R_{Q_j}(-\sigma_j\theta)OR_{P_j}(\theta) = O.
\end{equation}
We conclude that twirling over $\mathcal{J}_1$ therefore results in a
mapping from $\Pauli{XI}$ to $\Pauli{XZ}$ under a new noisy CZ gate
with an expected fidelity of $(f_\Pauli{XI} + f_\Pauli{YI})/2$.  The
same applies for $\Pauli{YI}$, and the additional twirling therefore
has the effect of averaging fidelities $f_\Pauli{XI}$ and
$f_\Pauli{YI}$.

Applying twirl gates on both qubits, that is, adding a second twirl
set $\mathcal{J}_2 = \{II, IZ\}$, does not change the above result,
but does result in the averaging of the fidelities for Pauli terms
$\Pauli{IX}$ and $\Pauli{IY}$, as well as for $\Pauli{XX}$,
$\Pauli{XY}$, $\Pauli{YX}$, and $\Pauli{YY}$, as shown in
Figure~\ref{Fig:CZGate}(c). More generally, we see from the
representation of class-3 gates in Figure~\ref{Fig:Classes} that
twirling using Pauli \Pauli{CI} and \Pauli{ID} rotations effectively
amounts to averaging the fidelities of corresponding terms in the
first two columns as well as those in the last two rows. The same
applies for Clifford operators in class 4.

The SWAP gate is an example of a class-2 operator where weight-one
Paulis supported on the first qubit all map to weight-one terms
supported on the second qubit, and vice versa due to Hermiticity. That
means that there are now three possible rotations $R_{P_j}$ we can
apply on the first qubit, each resulting in the averaging of two of
the Pauli terms. It may seem that nesting the twirls can help mix all
terms, but doing so generates a total of $2^3 = 8$ paths through the
noisy operator, which implies that the averaging cannot be uniform,
since $3$, the number of non-identity Pauli terms, does not divide
$8$, the number of elements in the twirl set. Instead, we can simply
set $\mathcal{J} = \{X_1, Y_1, Z_1\}$ and twirl over
$\{R_{\Pauli{X}_1}, R_{\Pauli{Y}_1}, R_{\Pauli{Z}_1}\}$, which maps
each single-qubit Pauli term to each of the three Pauli terms with
equal probability. A similar twirl can be applied on the second qubit.
The effect of applying this twirl on class-2 operators thus amounts to
averaging the fidelities of corresponding elements in the first three
columns and the last three rows in Figure~\ref{Fig:Classes}, thus
fully mixing the fidelities in the weight-two block.  For class 1 it
is possible to have one or three weight-one terms supported on the
first qubit to map to weight-one terms. In case there is only one we
average the fidelities in the other two columns, otherwise we can
average the elements in the first three columns. The same averaging
across rows applies for the weight-one terms in the right block,
depending on the number of terms that remain weight one following
conjugation. Since class-1 operators correspond to tensor products of
two single-qubit Clifford operations, it immediately follows that the
same techniques can be used for single-qubit gates.

In order to automatically determine the possible rotation twirls for a
given Hermitian two-qubit Clifford, we can first classify the operator
by conjugating all six weight-one Pauli terms. Depending on the
resulting weights, and the terms for which the weight changes we can
then easily determine the appropriate twirl.

Various twirl groups have been proposed in the literature, see for
example~\cite{CAR2015WEa,HAS2018FGWa,HEL2019XVWa}, and it is natural
to ask whether such twirls can further reduce the number of unique
Pauli fidelities through averaging. The short answer is that this is
indeed possible. However, implementing such twirls in the context of
two-qubit gates may no longer be possible using only single-qubit
gates. Inserting one or more two-qubit gates to twirl the noise
associated with a single two-qubit gate clearly introduces
disproportional levels of noise to the circuit and significantly
complicates learning. Similar difficulties arise when considering
twirling noise in the context of non-Clifford
gates~\cite{LAY2024MSa}.

\subsection{Extracting individual fidelities}\label{Sec:CorrectionSequences}

\begin{figure}
\centering
\begin{tabular}{ccc}
\includegraphics[width=0.35\textwidth]{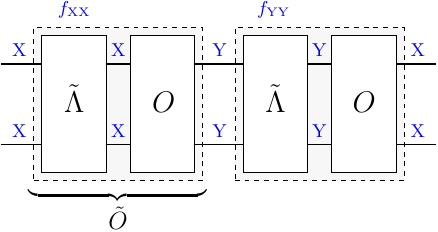}&
\hspace*{10pt}
&
\includegraphics[width=0.325\textwidth]{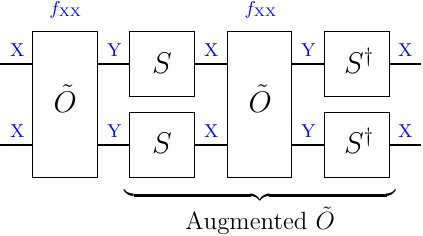}\\[2pt]
(a) && (b)
\end{tabular}
\caption{The noise channel of a Hermitian Clifford gate can be learned
  by repeated application of the gate. In (a) we trace the evolution a
  Pauli \Pauli{XX} operator through two instances of a noisy gate
  $\tilde{O}$ implementing a CZ gate affected by a Pauli noise channel
  $\tilde{\Lambda}$. Applying the noise channel on a Pauli operator
  incurs a multiplicative fidelity term corresponding to that Pauli,
  in that case resulting in an overall fidelity of
  $f_{\Pauli{XX}}f_{\Pauli{YY}}$. To learn $f_{\Pauli{XX}}$ up to a
  sign, we can (b) insert a correction sequence of single-qubit gates,
  mapping \Pauli{YY} back to \Pauli{XX} after each application of
  $\tilde{O}$. This process can equivalently be interpreted as
  augmenting the second $\tilde{O}$ with a fixed twirl
  sequence.}\label{Fig:LearningSequences}
\end{figure}

As explained in Section~\ref{Sec:NoiseModel}, we can learn the
fidelities of a Pauli noise channel associated with a Hermitian
Clifford operator by applying the noisy operator an even number of
times and taking measurements in different bases. Due to the presence
of the Clifford gate, those learning sequences typically only allow us
to learn pairwise products of certain fidelities. For instance, for a
noisy CZ gate, starting with a Pauli \Pauli{XX} term, we can learn the
fidelity pair $f_{\Pauli{XX}}f_{\Pauli{YY}}$, as illustrated in
Figure~\ref{Fig:LearningSequences}(a). In this case, application of
the CZ gate changes \Pauli{XX} to \Pauli{YY}, which preserves the
support and can be mapped back to the original Pauli by means of
single-qubit $R_Z(\pi/2)$ operators or, equivalently, phase gates (see
also~\cite{BER2023MKTa,CHE2020YZFa}).  The insertion of these
correction gates is shown in Figure~\ref{Fig:LearningSequences}(b)
along with the changes of Pauli terms and fidelities. Although such a
transformation is clearly always possible to correct terms that retain
their support, it is worth asking if doing so affects the learnability
of other terms in the same basis, for instance \Pauli{XI} or
\Pauli{IX}. Care needs to be taken as well not to introduce
inadvertent sign changes in some of the fidelity terms. The learning
sequence in Figure~\ref{Fig:LearningSequences}(b) can also be
interpreted as applying two instances of the twirled noisy gate, one
with a singleton twirl set
$\mathcal{J}_1 = \{(\Pauli{I}_1, \Pauli{I}_2)\}$, and one
$\mathcal{J}_2 = \{(\Pauli{Z}_1, \Pauli{Z}_2)\}$, where, with some
abuse of notation, we use tuples to indicate a pair of single-qubit
rotations. We see from Eq.~\eqref{Eq:C=RQCRP} that this construction
ensures that no spurious sign changes occur, but the question then
changes to whether or not we can always implement the desired
correction sequences using these singleton twirls.

Given an arbitrary basis \Pauli{ST} with
$\Pauli{S}, \Pauli{T} \in \{X,Y,Z\}$, our proposed approach is to
insert single-qubit correction gates whenever \Pauli{ST} maps to some
other weight-two Pauli. This clearly works for Clifford operators in
classes 1 and 2, since all weight-one terms retain their weight,
allowing us to insert any single-qubit rotation operator. For classes
3 and 4, it can be seen from Figure~\ref{Fig:Classes} that we only
ever need to exchange the terms in pairs $(\Pauli{A}, \Pauli{B})$ and
$(\Pauli{E}, \Pauli{F})$. This can be done respectively using Pauli
rotations \Pauli{CI} and \Pauli{ID}, both of which map to weight-one
Paulis and are therefore feasible candidates for the (nested) twirl
set. The lack of support changes for weight-one terms for operators in
class 1 then allows us to conclude that the proposed construction can
indeed generate all required correction sequences.

As seen from Figure~\ref{Fig:CZGate}(b), the insertion of correction
sequences can affect the fidelity pairs that arise from an initial
weight-one Pauli operator. Since the symmetry condition was assumed to
hold only for Pauli pairs $P_i$ and $Q_i = \sigma_i OP_iO^{\dag}$, we
can ask if the learning sequences arising from the proposed approach
still allow us to learn the original fidelity pair
$f_{P_i}f_{Q_i}$. In other words, can we still learn this fidelity
pair in one of the nine two-qubit Pauli bases?  To answer this
question, we consider the four different classes in turn.  The
question does not apply for operators in class 1, since all Pauli
terms maintain their original support.  For class 2, we can assume
that $\Pauli{AI}$ maps to $\Pauli{IF}$, which implies that
$\Pauli{AF}$ must map to itself. When working in the $\Pauli{AF}$
basis we therefore do not insert any correction gates, which means we
can measure the fidelity pair for $\Pauli{AI}$ and $\Pauli{IF}$ in
this basis. The same applies for all (weight-one) terms that change
support as well as for Paulis $\Pauli{CI}$ and $\Pauli{ID}$ in class
4.  What remains are the weight-two Pauli operators in classes 3 and 4
that map to weight-one Pauli operators, and vice versa.  Clearly,
since we only insert single-qubit correction gates for weight-two
Paulis that retain their weight, we do not insert any such gates here
and therefore always measure the original fidelity pairs in these
bases.  For class 4, we alternatively observe that for each row and
column in the weight-two box, there exists a weight-two Pauli that
maps to itself. For any Pauli $\Pauli{SI}$ or $\Pauli{IT}$ we can
therefore find a Pauli basis in the same row or column for which the
circuit is run without correction gates, and for which we therefore
measure the original fidelity pairs. We conclude that we can measure
the original fidelity pairs in at least one basis for Paulis that
change their support.

\section{Basis selection using graph coloring and covering arrays}\label{Sec:BasisSelection}

We now consider the selection of state preparation and measurement
Pauli bases that allow us to measure all fidelities for our benchmark
set $\mathcal{B}$, which then provides all the information needed to
determine the sparse Pauli-Lindblad noise model parameters. To better
explain the intuition behind our methods for basis selection, we begin
with a simple setting. We consider a noisy layer of identity gates,
for which fidelities can be measured individually rather than in pairs
and equate $\mathcal{B}$ to the set of terms $\mathcal{K}$ for a
two-local noise model. We then successively discuss the changes and
considerations that apply for more general settings.

\subsection{Basis selection for two-local models}
In the most basic setting of a two-local noise model,
Theorem~\ref{Thm:FullRankM} guarantees that it suffices to measure the
fidelities of all weight-one Paulis on the model qubits, and all
weight-two Paulis on neighboring qubits. For this it suffices to
select a series of bases such that all nine
$\{X,Y,Z\}\times \{X,Y,Z\}$ weight-two Pauli terms occur at least once
for each connected pair of qubits.

We now leverage the combinatorial construction called \emph{covering
  array} for basis selection.  The covering array CA$(N; t, k, v)$ is
a $N$-by-$k$ matrix with elements from an alphabet of size $v$, such
that each $N$-by-$t$ submatrix contains each of the $v^t$ possible
$t$-tuples of symbols in at least one row~\cite{COL2004a}. The
covering array number CAN$(t,k,v)$ is the smallest $N$ for which
CA$(N; t, k, v)$ exists. In order to use these covering arrays, we
start the basis-selection protocol by generating a graph with vertices
corresponding to the qubits in the noise model and edges for the
neighboring qubits, which can include virtual connections that
indicate expected crosstalk terms between the given pairs of
qubits. Once the graph has been generated we apply graph coloring and
find a covering array CA$(N; 2, k, 3)$ of strength $t=2$, where $k$
matches the number of colors and the symbols of the alphabet are set
to $\{X,Y,Z\}$ (for more information on covering arrays of order three
with strength two, see~\cite{TOR2021AAa}). By mapping each of the
colors to a column in the covering array we can then construct $N$
bases.  Each basis is generated by a single row of the array, which
prescribes the mapping of a color to a Pauli term. The Pauli basis
string can then be formed by looking up the terms associated with the
color assigned to qubits in the graph coloring. We summarize the
different steps in Figure~\ref{Fig:BasisSelection}.

An equivalent approach for basis selection for two-local models was
proposed independently in~\cite{arXiv:2311.11639}, but without making
the connection to covering arrays. Perfect hash families, which are
related to covering arrays~\cite{COL2024a}, were used for basis
selection in~\cite{COT2020Wa} to measure all $k$-local Pauli
observables. A randomized approach to basis selection in this context,
aimed at reducing the number of measurements rather than the number of
unique bases, was proposed in~\cite{EVA2019HFa} and derandomized
in~\cite{HUA2021KPa}.
 
\subsection{Reduction of the basis elements}
When noise occurs in the context of general layers of Hermitian
Clifford gates, noise channel fidelities typically occur as the
product $f_if_j$ of the fidelities corresponding to Paulis $P_i$ and
$P_j$. For convenience we refer to $P_i$ and $P_j$ as the Pauli pair
or Pauli terms that appear in the fidelity, and $f_if_j$ the fidelity
or fidelity pair for the Pauli pair. Although it is not always
possible to uniquely determine the individual fidelities, there are
some cases, as discussed in Section~\ref{Sec:CorrectionSequences},
where it is nevertheless possible to do so by inserting correction
sequences in the learning circuits. These sequences are aimed at
correcting Pauli terms that maintain their support following
conjugation by the gates in the layer, but one side effect of these
sequences is that they also affect the terms that appear in other
fidelity pairs. If these terms change support we may require
strengthening the symmetry condition to apply to all such pairs as
well. We partially addressed this in
Section~\ref{Sec:CorrectionSequences}, where we showed that for
one-local terms $P$ there is always at least one basis in which we can
learn the original fidelity pair, which would then allow resolution of
the individual fidelities as a consequence of the symmetry
condition. This result does not generally extend to Pauli terms that
overlap with multiple two-qubit gates. For instance, consider a
weight-two Pauli string $P_1P_2$ such that each term $P_i$ is
supported on a different two-qubit gate. If application of the gates
does not change the support, and therefore maps the Pauli to some
weight-two Pauli $Q_1Q_2$, then for each $i\in\{1,2\}$ there is at
least one basis for gate $i$ for which Pauli $P_i$ maps to the
original term $Q_i$. However, since the bases for the gates may be
chosen independently, there is no guarantee that we select a pair of
bases for which the original Pauli pair appears simultaneously, that
is, for which $P_1P_2$ maps to $Q_1Q_2$. Those cases either require
additional learning bases or stronger symmetry assumptions.

\begin{figure}[!t]
\centering
\includegraphics[width=0.99\textwidth]{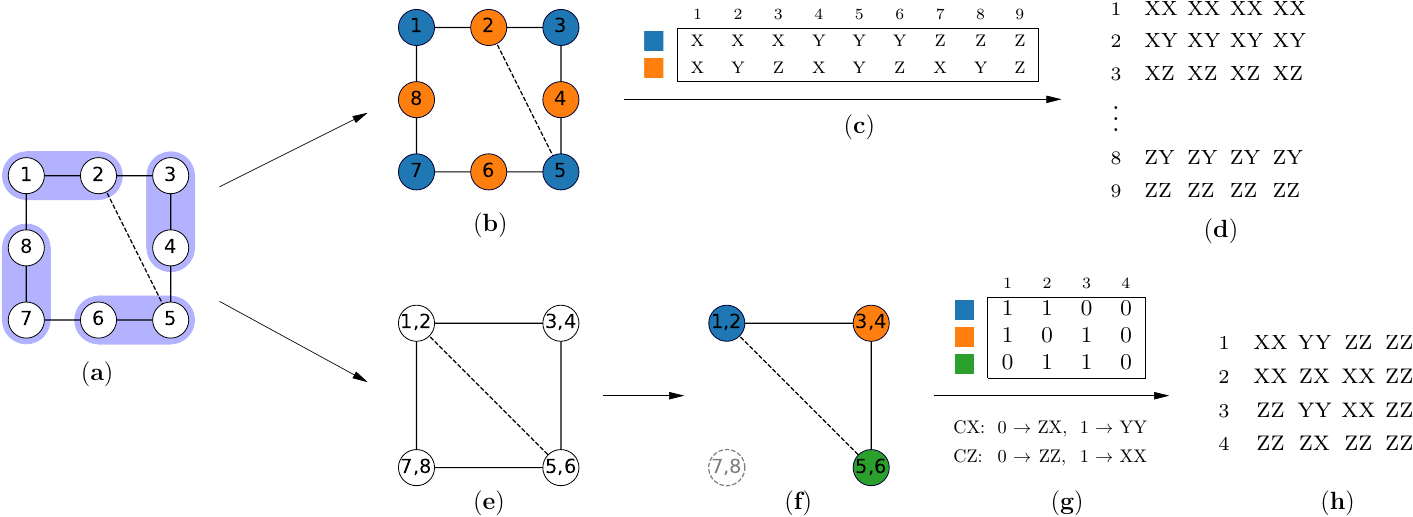}
\caption{Application of the learning protocol to (a) a layer of gates
  $\mbox{CZ}(1,2)$, $\mbox{CX}(3,4)$, $\mbox{CZ}(5,6)$, and
  $\mbox{Swap}(7,8)$ on an example topology with presumed crosstalk
  between qubits 2 and 5.  The top row shows basis selection with
  regular Pauli twirling. Here, we start by (b) finding a graph
  coloring for the qubit topology. We then find (c) a covering array
  CA(2,2,3) with symbols $\{X,Y,Z\}$, here shown in transposed
  form. Each column identifies the mapping of colors to symbols for a
  single basis. Applying the mapping gives (d) the desired learning
  bases.  The bottom row illustrates basis selection when using a
  Pauli rotation twirl in addition to the regular Pauli twirl. In this
  case, we first (e) merge the vertices of the qubits of each gate,
  and, depending on the type of gate, associate with each vertex one
  or two symbols. We can then (f) delete all vertices that require
  only a single symbol, for instance for the swap gate, since any
  basis terms on those qubits allow us to learn all desired
  fidelities. We then color the resulting graph and (g) find a
  CA(2,3,2) covering array and define a gate-dependent mapping of
  symbols to basis terms. Finally, we (h) apply the mapping to obtain
  the different bases. Note that the chromatic number of the graphs
  corresponding to basic and rotation twirling can differ; in the
  present example the chromatic numbers are two and three,
  respectively. However, if the additional crosstalk occurred on
  qubits 2--4, rather than 2--5, the chromatic numbers would change to
  three and two, respectively.  }\label{Fig:BasisSelection}
\end{figure}

Combining Pauli twirls with the rotational twirls, introduced in
Section~\ref{Sec:ExtraTwirling}, allows us to do away with correction
sequences altogether: rather than correcting the Pauli terms that
appear in each fidelity pair, we shape the noise channel such that
certain fidelities are averaged and become equivalent. Whenever
several fidelities are equivalent it suffices to measure any one of
them to know them all, which allows us to reduce the number of
measurement bases. Consider for example the classes of two-qubit gates
in Figure~\ref{Fig:Classes}. For classes 1 and 2, the fidelities for
all Paulis with the same support are averaged. It therefore suffices
to take measurements in, say, the CD basis (where $C$ and $D$ are
mapped to appropriate Pauli terms), which includes terms for all three
support $CI$, $ID$, and $CD$. The rotation twirl set for classes 3 and
4 is smaller and therefore leads to averaging of smaller subsets of
fidelities. Nevertheless, even under the weak symmetry assumption, we
can still measure all desired fidelities by measuring for instance
only in the CD and AF bases. In this case, measuring the fidelities
for Pauli terms that are supported on multiple gates does not require
any special treatment. The implication of these simplifications is
that we can merge any pair of vertices in the graph whose qubits are
subject to the same two-qubit gate. The number of basis elements or
symbols required for the resulting vertex is one for gates from
classes 1 or 2, and two for classes 3 and 4. Vertices corresponding to
idle qubits or qubits on which single-qubit operations apply also
require only a single symbol. After merging the vertices, we can
delete all vertices with a single symbol because the corresponding
term in the learning basis can be arbitrarily set to any
Pauli. Following these transformations we can apply graph coloring and
determine the bases using a (binary) covering array CA$(2,k',2)$,
where the number of colors $k'$ can differ from that of the original
graph, and where the alphabet size $v$ is now equal to 2 (explicit
constructions for binary covering arrays can be found
in~\cite{TOR2012Ra}). It can be seen from Table~\ref{Table:CAN}, that
the smaller alphabet size leads to a substantial reduction in the
number of learning bases. When extracting the learning bases from the
graph coloring and the binary covering array, we expand each symbol
for merged vertices to two gate-dependent Pauli terms, one for each
qubit associated with the original vertices. We illustrate the
basis-selection protocol in the case of Pauli twirls, with and without
additional rotation twirls in Figure~\ref{Fig:BasisSelection}

\begin{table}
  \centering\footnotesize\setlength{\tabcolsep}{5.5pt}
\begin{tabular}{|l|rrrrrrrrrrrrrrrrrrr|}
\hline
\multicolumn{1}{|r|}{$k$} & 2 & 3 & 4 & 5 & 6 & 7 & 8 & 9 & 10
 & 11 & 12 & 13 & 14 & 15 & 16 & 17 & 18 & 19 & 20\\
\hline
CAN$(2,k,2)$ & 4 & 4 & 5 & 6 & 6 & 6 & 6 & 6 & 6
 & 7&7&7&7&7 &8 &8 &8 &8 &8\\
CAN$(3,k,2)$ & -- &8&8&10&12&12&12&12&12
&12&15&16&16 &17&17 & 18&18&18&18\\
CAN$(4,k,2)$ & -- & -- & 16&16 &21 &24&24&24&24
&24&24 & 32 & 35&35&35&35 & 36&39&39\\
\hline
CAN$(2,k,3)$ & 9 & 9 & 9 & 11&12&12 &13&13 & 14
& 15&15&15&15&15 & 15&15&15&15&15\\
CAN$(3,k,3)$ & -- & 27&27 & 33&33& 39 &42&45&45
& 45&45&45&45& 51&51 & 58 & 59&59&59\\
\hline
\end{tabular}
\caption{Number of rows (learning bases) for the smallest
  covering arrays of different sizes known at the time of
  writing~\cite{COL-CoveringArrayTables,TOR-CoveringArrayTables}. The
  values for CAN$(2,k,2)$ match the R\'{e}nyi lower
  bound~\cite{REN1971a} and are therefore optimal.}\label{Table:CAN}
\end{table}

\subsection{Basis selection for \texorpdfstring{$\ell$}{l}-local noise models}
We now show how the learning protocol can be modified to support the
generalization from the two-local Pauli-Lindblad noise model to any
weight $\ell$-local noise models. We can define an $\ell$-local noise
model by selecting support sets $\{\mathcal{S}_i\}_i$ defined in
Theorem~\ref{Thm:FullRankM} with
$\max_i \vert\mathcal{S}_i\vert = \ell$ and setting
$\mathcal{K} = \bigcup_i \mathcal{P}_i$, where $\mathcal{P}_i$ is the
set of all Pauli terms supported on $\mathcal{S}_i$.  In order to
learn the model parameters is then suffices to learn the fidelities
for Pauli terms in $\mathcal{B} := \mathcal{K}$. For this we again set
up a graph with vertices corresponding to the model qubits. The edges
are generated based on each set $\mathcal{S}_i$ such that the graph
has a clique on the vertices associated with the qubits in
$\mathcal{S}_i$. After this, depending on the twirl group, we can
merge the vertices of the qubits in each of the gates and color the
graph. Observe that addition of the local cliques ensures that, for
each $i$, all vertices in $\mathcal{S}_i$ are assigned different
colors. The basis selection is then done using a covering array
CA$(\ell,k,v)$, where $k$ is the number of colors in the (reduced)
graph, and $v$ is 2 or 3 depending on whether or not we apply
rotational twirling. Mapping the symbols back to Pauli terms proceeds
as before.

\subsection{Further extensions}
Finally, we briefly consider basis selection for general sets of Pauli
terms $\mathcal{B}$ and an isolated noise channel. It is well known
that sets of mutually commuting Paulis observables can be measured
simultaneously by means of a basis change effected by an appropriate
Clifford circuit. In this case, we can determine the minimum number of
learning bases by forming a graph with vertices corresponding to the
elements in $\mathcal{B}$ and edges between vertices whose Pauli terms
commute and then finding the maximum clique cover. Alternatively, we
can connect Pauli terms that do not commute and find a graph
coloring. Each of the cliques in the cover, or terms with equal cover,
can then be measured simultaneously under the appropriate basis
transformation. However, this approach is generally not practically
feasible on noisy quantum processors due to the excessive noise
generated by the Clifford circuit implementing the basis change. This
problem can be overcome, possibly at the expense of more bases, by
allowing Pauli operators to be measured simultaneously only if the
Pauli terms on each of the qubits commute. When considering learning
of noise in the context of gates a more elaborate approach is needed
that takes advantage of the fact that, under the symmetry assumption,
we can learn the fidelities for Pauli pair $P_i$ and $P_j$ by
measuring either of the two Pauli operators. The equivalence of
fidelities due to rotational twirling further increases the number of
Pauli observables that can be measured to obtain the desired
fidelities. We will not discuss any of these extensions here.

\section{Conclusions}\label{Sec:Conclusions}

We have shown how the combined use of graph coloring and covering
arrays gives rise to an elegant protocol for basis selection in the
learning of sparse Pauli-Lindblad noise models. Complementary to this
we have introduced Pauli-rotation twirls and shown how they can be
used to reduce the number of unique fidelities in Pauli noise
channels. Together, these techniques enable the practical learning of
noise models for highly-connected qubit topologies with added
connections to model crosstalk between known qubit pairs. Based on the
the covering array sizes listed in Table~\ref{Table:CAN}, we can learn
two-local noise model for Pauli-twirled channels even when the qubit
topology has a chromatic number up to twenty by leveraging the
covering array in~\cite{NUR2004a}. Adding a rotation twirl allows us
to reduce the alphabet size of the covering arrays, which drastically
reduces the required number of learning bases and opens up the avenue
to learning three-local noise models. Reduction of the number of
unique Pauli fidelities by means of the rotation twirl is generally
applicable and could be used, for instance, to reduce the number of
measurements required for averaged circuit eigenvalue
sampling~\cite{FLA2022a,HOC2024DHa}.

\section*{Acknowledgements}

EvdB would like to thank Dr.~Jose Torres-Jimenez for providing access
to the uniform covering arrays on~\cite{TOR-CoveringArrayTables} and
for pointing out useful references. We would also like to thank the
referees for providing additional pointers to relevant literature.

\bibliographystyle{quantum}
\bibliography{bibliography}

\begin{thebibliography}{10}

\bibitem{PER2014MSYa}
Alberto Peruzzo, Jarrod McClean, Peter Shadbolt, Man-Hong Yung, Xiao-Qi Zhou,
  Peter~J. Love, Al\'{a}n Aspuru-Guzik, and Jeremy~L. O’Brien.
\newblock ``A variational eigenvalue solver on a photonic quantum processor''.
\newblock \href{https://dx.doi.org/10.1038/ncomms5213}{Nature Communications
  {\bf 5}, 4213}~(2014).

\bibitem{MCC2016RBAa}
Jarrod~R. McClean, Jonathan Romero, Ryan Babbush, and Al{\'a}n Aspuru-Guzik.
\newblock ``The theory of variational hybrid quantum-classical algorithms''.
\newblock \href{https://dx.doi.org/10.1088/1367-2630/18/2/023023}{New Journal
  of Physics {\bf 18}, 023023}~(2016).

\bibitem{TIL2022CCPa}
Jules Tilly, Hongxiang Chen, Shuxiang Cao, Dario Picozzi, Kanav Setia, Ying Li,
  Edward Grant, Leonard Wossnig, Ivan Rungger, George~H Booth, et~al.
\newblock ``The variational quantum eigensolver: a review of methods and best
  practices''.
\newblock \href{https://dx.doi.org/10.1016/j.physrep.2022.08.003}{Physics
  Reports {\bf 986}, 1--128}~(2022).

\bibitem{BER2022MTa}
Ewout van~den Berg, Zlatko~K. Minev, and Kristan Temme.
\newblock ``Model-free readout-error mitigation for quantum expectation
  values''.
\newblock \href{https://dx.doi.org/10.1103/PhysRevA.105.032620}{Physical Review
  A {\bf 105}, 032620}~(2022).

\bibitem{BRA2021SKMa}
Sergey Bravyi, Sarah Sheldon, Abhinav Kandala, David McKay, and Jay~M.
  Gambetta.
\newblock ``Mitigating measurement errors in multiqubit experiments''.
\newblock \href{https://dx.doi.org/10.1103/PhysRevA.103.042605}{Physical Review
  A {\bf 103}, 042605}~(2021).

\bibitem{HIC2021BNa}
Rebecca Hicks, Christian~W. Bauer, and Benjamin Nachman.
\newblock ``Readout rebalancing for near-term quantum computers''.
\newblock \href{https://dx.doi.org/10.1103/PhysRevA.103.022407}{Phys. Rev. A
  {\bf 103}, 022407}~(2021).

\bibitem{TEM2017BGa}
Kristan Temme, Sergey Bravyi, and Jay~M. Gambetta.
\newblock ``Error mitigation for short-depth quantum circuits''.
\newblock \href{https://dx.doi.org/10.1103/PhysRevLett.119.180509}{Physical
  Review Letters {\bf 119}, 180509}~(2017).

\bibitem{LI2017Ba}
Ying Li and Simon~C. Benjamin.
\newblock ``Efficient variational quantum simulator incorporating active error
  minimization''.
\newblock \href{https://dx.doi.org/10.1103/PhysRevX.7.021050}{Physical Review X
  {\bf 7}, 021050}~(2017).

\bibitem{KAN2019TCMa}
Abhinav Kandala, Kristan Temme, Antonio~D. C{\'o}rcoles, Antonio Mezzacapo,
  Jerry~M. Chow, and Jay~M. Gambetta.
\newblock ``Error mitigation extends the computational reach of a noisy quantum
  processor''.
\newblock \href{https://dx.doi.org/10.1038/s41586-019-1040-7}{Nature {\bf 567},
  491--495}~(2019).

\bibitem{BER2023MKTa}
Ewout van~den Berg, Zlatko~K. Minev, Abhinav Kandala, and Kristan Temme.
\newblock ``Probabilistic error cancellation with sparse {P}auli-{L}indblad
  models on noisy quantum processors''.
\newblock \href{https://dx.doi.org/10.1038/s41567-023-02042-2}{Nature Physics
  {\bf 19}, 1116--1121}~(2023).

\bibitem{END2018BLa}
Suguru Endo, Simon~C. Benjamin, and Ying Li.
\newblock ``Practical quantum error mitigation for near-future applications''.
\newblock \href{https://dx.doi.org/10.1103/PhysRevX.8.031027}{Physical Review X
  {\bf 8}, 031027}~(2018).

\bibitem{MAR2021SZa}
Andrea Mari, Nathan Shammah, and William~J Zeng.
\newblock ``Extending quantum probabilistic error cancellation by noise
  scaling''.
\newblock \href{https://dx.doi.org/10.1103/PhysRevA.104.052607}{Physical Review
  A {\bf 104}, 052607}~(2021).

\bibitem{FER2024HVNa}
Samuele Ferracin, Akel Hashim, Jean-Loup Ville, Ravi Naik, Arnaud
  Carignan-Dugas, Hammam Qassim, Alexis Morvan, David~I. Santiago, Irfan
  Siddiqi, and Joel~J. Wallman.
\newblock ``Efficiently improving the performance of noisy quantum computers''.
\newblock \href{https://dx.doi.org/10.22331/q-2024-07-15-1410}{{Quantum} {\bf
  8}, 1410}~(2024).

\bibitem{MCD2022MSSa}
Benjamin McDonough, Andrea Mari, Nathan Shammah, Nathaniel~T. Stemen, Misty
  Wahl, William~J. Zeng, and Peter~P. Orth.
\newblock ``Automated quantum error mitigation based on probabilistic error
  reduction''.
\newblock In 2022 IEEE/ACM Third International Workshop on Quantum Computing
  Software (QCS).
\newblock \href{https://dx.doi.org/10.1109/QCS56647.2022.00015}{Pages 83--93}.
\newblock IEEE~(2022).

\bibitem{KIM2023EAWa}
Youngseok Kim, Andrew Eddins, Sajant Anand, Ken~Xuan Wei, Ewout van~den Berg,
  Sami Rosenblatt, Hasan Nayfeh, Yantao Wu, Michael Zaletel, Kristan Temme, and
  Abhinav Kandala.
\newblock ``Evidence for the utility of quantum computing before fault
  tolerance''.
\newblock \href{https://dx.doi.org/10.1038/s41586-023-06096-3}{Nature {\bf
  618}, 500--505}~(2023).

\bibitem{PhysRevLett.76.722}
Charles~H. Bennett, Gilles Brassard, Sandu Popescu, Benjamin Schumacher,
  John~A. Smolin, and William~K. Wootters.
\newblock ``Purification of noisy entanglement and faithful teleportation via
  noisy channels''.
\newblock \href{https://dx.doi.org/10.1103/PhysRevLett.76.722}{Phys. Rev. Lett.
  {\bf 76}, 722--725}~(1996).

\bibitem{knill2004fault}
Emanuel Knill.
\newblock ``Fault-tolerant postselected quantum computation: Threshold
  analysis''~(2004)
  \href{http://arxiv.org/abs/quant-ph/0404104}{arXiv:quant-ph/0404104}.

\bibitem{kern2005quantum}
Oliver Kern, Gernot Alber, and Dima~L. Shepelyansky.
\newblock ``Quantum error correction of coherent errors by randomization''.
\newblock \href{https://dx.doi.org/10.1140/epjd/e2004-00196-9}{The European
  Physical Journal D-Atomic, Molecular, Optical and Plasma Physics {\bf 32},
  153--156}~(2005).

\bibitem{geller2013efficient}
Michael~R Geller and Zhongyuan Zhou.
\newblock ``Efficient error models for fault-tolerant architectures and the
  {P}auli twirling approximation''.
\newblock \href{https://dx.doi.org/10.1103/PhysRevA.88.012314}{Physical Review
  A {\bf 88}, 012314}~(2013).

\bibitem{wallman2016noise}
Joel~J. Wallman and Joseph Emerson.
\newblock ``Noise tailoring for scalable quantum computation via randomized
  compiling''.
\newblock \href{https://dx.doi.org/10.1103/PhysRevA.94.052325}{Physical Review
  A {\bf 94}, 052325}~(2016).

\bibitem{FLA2020Wa}
Steven~T. Flammia and Joel~J. Wallman.
\newblock ``Efficient estimation of {P}auli channels''.
\newblock \href{https://dx.doi.org/10.1145/3408039}{ACM Transactions on Quantum
  Computing {\bf 1}, 1--32}~(2020).

\bibitem{ERH2019WPMa}
Alexander Erhard, Joel~J. Wallman, Lukas Postler, Michael Meth, Roman Stricker,
  Esteban~A. Martinez, Philipp Schindler, Thomas Monz, Joseph Emerson, and
  Rainer Blatt.
\newblock ``Characterizing large-scale quantum computers via cycle
  benchmarking''.
\newblock \href{https://dx.doi.org/10.1038/s41467-019-13068-7}{Nature
  Communications {\bf 10}, 1--7}~(2019).

\bibitem{PhysRevX.4.011050}
Shelby Kimmel, Marcus~P. da~Silva, Colm~A. Ryan, Blake~R. Johnson, and Thomas
  Ohki.
\newblock ``Robust extraction of tomographic information via randomized
  benchmarking''.
\newblock \href{https://dx.doi.org/10.1103/PhysRevX.4.011050}{Phys. Rev. X {\bf
  4}, 011050}~(2014).

\bibitem{HEL2019XVWa}
Jonas Helsen, Xiao Xue, Lieven M.~K. Vandersypen, and Stephanie Wehner.
\newblock ``A new class of efficient randomized benchmarking protocols''.
\newblock \href{https://dx.doi.org/10.1038/s41534-019-0182-7}{npj Quantum
  Information {\bf 5}, 1--9}~(2019).

\bibitem{CHE2020YZFa}
Senrui Chen, Yunchao Liu, Matthew Otten, Alireza Seif, Bill Fefferman, and
  Liang Jiang.
\newblock ``The learnability of {P}auli noise''.
\newblock \href{https://dx.doi.org/10.1038/s41467-022-35759-4}{Nature
  Communications {\bf 14}, 52}~(2023).

\bibitem{cai2019constructing}
Zhenyu Cai and Simon~C Benjamin.
\newblock ``Constructing smaller {P}auli twirling sets for arbitrary error
  channels''.
\newblock \href{https://dx.doi.org/10.1038/s41598-019-46722-7}{Scientific
  reports {\bf 9}, 1--11}~(2019).

\bibitem{CAR2015WEa}
Arnaud Carignan-Dugas, Joel~J. Wallman, and Joseph Emerson.
\newblock ``Characterizing universal gate sets via dihedral benchmarking''.
\newblock \href{https://dx.doi.org/10.1103/PhysRevA.92.060302}{Phys. Rev. A
  {\bf 92}, 060302}~(2015).

\bibitem{HAS2018FGWa}
A.~K. Hashagen, S.~T. Flammia, D.~Gross, and J.~J. Wallman.
\newblock ``Real randomized benchmarking''.
\newblock \href{https://dx.doi.org/10.22331/q-2018-08-22-85}{{Quantum} {\bf 2},
  85}~(2018).

\bibitem{LAY2024MSa}
David Layden, Bradley Mitchell, and Karthik Siva.
\newblock ``Theory of quantum error mitigation for non-{C}lifford
  gates''~(2024).
\newblock  \href{http://arxiv.org/abs/2403.18793}{arXiv:2403.18793}.

\bibitem{COL2004a}
Charles~J. Colbourn.
\newblock ``Combinatorial aspects of covering arrays''.
\newblock Le Matematiche {\bf 59}, 125--172~(2004).
\newblock
  url:~\url{https://lematematiche.dmi.unict.it/index.php/lematematiche/article/view/166}.

\bibitem{TOR2021AAa}
Jose Torres-Jimenez, Brenda Acevedo-Ju{\'a}rez, and Himer Avila-George.
\newblock ``Covering array {EXtender}''.
\newblock \href{https://dx.doi.org/10.1016/j.amc.2021.126122}{Applied
  Mathematics and Computation {\bf 402}, 126122}~(2021).

\bibitem{arXiv:2311.11639}
Jose~Este Jaloveckas, Minh Tham~Pham Nguyen, Lilly Palackal, Jeanette~Miriam
  Lorenz, and Hans Ehm.
\newblock ``Efficient learning of sparse {P}auli {L}indblad models for fully
  connected qubit topology''~(2023).
\newblock  \href{http://arxiv.org/abs/2311.11639}{arXiv:2311.11639}.

\bibitem{COL2024a}
Charles Colbourn.
\newblock ``Covering perfect hash families and covering arrays of higher
  index''.
\newblock
  \href{https://dx.doi.org/10.22108/ijgt.2023.137230.1836}{International
  Journal of Group Theory {\bf 13}, 293--305}~(2024).

\bibitem{COT2020Wa}
Jordan Cotler and Frank Wilczek.
\newblock ``Quantum overlapping tomography''.
\newblock \href{https://dx.doi.org/10.1103/PhysRevLett.124.100401}{Phys. Rev.
  Lett. {\bf 124}, 100401}~(2020).

\bibitem{EVA2019HFa}
Tim~J. Evans, Robin Harper, and Steven~T. Flammia.
\newblock ``Scalable {B}ayesian {H}amiltonian learning''~(2019).
\newblock  \href{http://arxiv.org/abs/1912.07636}{arXiv:1912.07636}.

\bibitem{HUA2021KPa}
Hsin-Yuan Huang, Richard Kueng, and John Preskill.
\newblock ``Efficient estimation of {P}auli observables by derandomization''.
\newblock \href{https://dx.doi.org/10.1103/PhysRevLett.127.030503}{Phys. Rev.
  Lett. {\bf 127}, 030503}~(2021).

\bibitem{TOR2012Ra}
Jose Torres-Jimenez and Eduardo Rodriguez-Tello.
\newblock ``New bounds for binary covering arrays using simulated annealing''.
\newblock
  \href{https://dx.doi.org/https://doi.org/10.1016/j.ins.2011.09.020}{Information
  Sciences {\bf 185}, 137--152}~(2012).

\bibitem{COL-CoveringArrayTables}
Charles Colbourn.
\newblock \url{https://www.public.asu.edu/~ccolbou/src/tabby/catable.html}.

\bibitem{TOR-CoveringArrayTables}
Jose Torres-Jimenez.
\newblock \url{https://www.tamps.cinvestav.mx/~oc/}.

\bibitem{REN1971a}
Alfr\'{e}d R\'{e}nyi.
\newblock ``Foundations of probability''.
\newblock \href{https://dx.doi.org/}{Wiley}. New York, USA~(1971).

\bibitem{NUR2004a}
Kari~J. Nurmela.
\newblock ``Upper bounds for covering arrays by tabu search''.
\newblock
  \href{https://dx.doi.org/https://doi.org/10.1016/S0166-218X(03)00291-9}{Discrete
  applied mathematics {\bf 138}, 143--152}~(2004).

\bibitem{FLA2022a}
Steven~T. Flammia.
\newblock ``{Averaged Circuit Eigenvalue Sampling}''.
\newblock In Fran\c{c}ois Le~Gall and Tomoyuki Morimae, editors, 17th
  Conference on the Theory of Quantum Computation, Communication and
  Cryptography (TQC 2022).
\newblock \href{https://dx.doi.org/10.4230/LIPIcs.TQC.2022.4}{Volume 232 of
  Leibniz International Proceedings in Informatics (LIPIcs), pages 4:1--4:10}.
\newblock Dagstuhl, Germany~(2022). Schloss Dagstuhl -- Leibniz-Zentrum f{\"u}r
  Informatik.

\bibitem{HOC2024DHa}
Evan~T. Hockings, Andrew~C. Doherty, and Robin Harper.
\newblock ``Scalable noise characterisation of syndrome extraction circuits
  with averaged circuit eigenvalue sampling''~(2024).
\newblock  \href{http://arxiv.org/abs/2404.06545}{arXiv:2404.06545}.

\end{thebibliography}

\end{document}